\documentclass[preprint]{iopjournal}

\usepackage{booktabs}
\begin{document}


\title{How Physics Professors Use and Frame Generative AI Tools}

\author{Vidar Skogvoll$^{1,*}$\orcid{0000-0003-4941-6886} and Tor Ole Odden$^{1}$\orcid{0000-0003-1635-9491}}

\affil{$^1$Center for Computing in Science Education, Department of Physics, University of Oslo, Oslo, Norway}

\affil{$^*$Author to whom any correspondence should be addressed.}

\email{vidarsko@uio.no}


\begin{abstract}
Generative AI is rapidly reshaping how physicists teach, learn, and conduct research, yet little is known about how physics faculty are responding to these changes. 
We interviewed 12 physics professors at a major Scandinavian research university to explore their uses and perceptions of Generative AI (GenAI) in both teaching and research. 
Using the theoretical framework of epistemic framing, we conducted a thematic analysis that identified 19 overlapping practices, ranging from coding and literature review to assessment and feedback. 
From these practices, we derived six overlapping epistemic frames through which professors make sense of GenAI: as a threat to genuine learning and assessment, a source of knowledge, a discussion partner, a text-processing tool, a coding tool, and a labor-saving device. 
While the latter five position GenAI as a useful tool in the physicists' toolbox, the threat frame represented an overarching concern that colored all other frames. 
These findings reveal how GenAI is beginning to transform what it means to be a physicist, highlighting both opportunities for innovation and challenges for academic integrity and learning.
\end{abstract}

\section{Introduction}
Physics is a field that has been deeply shaped by technology, especially computers. 
100 years ago, physicists would spend considerable time finding ways to solve integrals analytically and looking up answers to undetermined integrals in books and tables. 
Data had to be written down, graphs had to be created by hand, literature was available through libraries, and all meetings were held in-person.  
Today, while some physicists still search for such analytical solutions, most rely on computers to deal with messy calculations and involved integrals. 
Computers now store our data, visualize our results, give us immediate access to unimaginable amounts of literature, and let us chat with colleagues continents away.

We are at the cusp of a new such revolution. 
Generative AI (GenAI) tools in the form of large language models now allow physicists to delegate even more of their cognitive work to computers. 
Instead of having to write code by hand, a physicist can now ask an AI assistant to write it for them. 
GenAI tools can summarize articles, translate documents, produce detailed research summaries and literature reviews, and even perform mathematical and numerical analyses. 
Whereas before, physicists needed to spend months or years learning to program in the language of the computer, now, for the first time, computers fluently speak \textit{our} language.

This revolution is not without challenges. 
These developments have shaken the foundations of academia, especially in science and science education \cite{erduran_ai_2023,Erduran2024Theimpactofartificialintellige}. 
GenAI tools can convincingly write papers, solve exam questions, and create lab reports, often of higher quality than the average student. 
From the point of view of academic integrity, GenAI challenges what authorship of a text really means, which has serious implications for a sector within which texts are the currency for communicating ideas, research, and results \cite{Yeadon2024EvaluatingAIandhumanauthorship, kortemeyer_cheat_2024}. 
Furthermore, from an educational standpoint, such tools raise critical questions as to how GenAI affects our cognitive development \cite{kosmyna_your_2025}.

Physics and physicists find themselves in a special place within this revolution. 
In contrast to fields like the humanities in which research is evaluated primarily based on the quality of literature and argumentation, physics remains primarily an empirical subject where experimental results, mathematical derivations, and numerical models are king. 
Thus, GenAI is likely to serve a different role for physics, potentially acting as an additional tool for many of the things physicists already do. 
Furthermore, physicists are generally tech-savvy, which may uniquely position them as informed users of these kinds of tools. 
An education in physics requires both facility with mathematical models and experience with different forms of computational technology, and GenAI is at its heart a computationally-run mathematical model of language. 

For all of these reasons, it is critically important to study how GenAI tools are affecting physics research and education. 
However, because of the newness of these tools, research in this area remains sparse.

In this study, we seek to understand how professionals in physics, i.e., physics professors, view generative AI tools and how they use them in both their research and teaching work. Specifically, we ask the following research questions:
\begin{enumerate}
    \item How are physics professors using GenAI tools in their teaching and research?
    \item How do physics professors frame these tools?
\end{enumerate}
In order to answer these questions, we have interviewed 12 physics professors at a major Scandinavian research university, asking them about the different ways they have taken up these tools in their teaching and research. 
We have inductively thematically analyzed these interviews based on the theoretical framework of epistemic framing \cite{Tannen1993}. 
This analysis has uncovered 19 professional and pedagogical practices, which we have organized into 6 overlapping epistemic frames. 

The article is structured as follows: 
In section 2, we briefly summarize the state of physics education research on GenAI in order to position the present work. 
Next, in section 3, we will present the theoretical framework used to analyze the interviews: epistemic framing. 
Thereafter, we present the details of our data collection and analysis method in section 4, before presenting the results of the analysis in section 5. 
We conclude with remarks and implications in section 6, commenting on how GenAI has begun to redefine what it means to be a physicist.

\section{Literature Review}

Generative AI has technically been around for many decades, but only recently seen an explosion in usage since the public release of ChatGPT in November of 2022. 
This explosion was made possible by ground-breaking work on large language models a few years prior when the \textit{transformer} architecture was introduced \cite{vaswaniAttentionAllYou2017}.
Today, these GenAI tools are general-purpose and multimodal agents, useful for brainstorming, text generation, discussion, problem-solving, coding, and more. 
Popular GenAI tools include ChatGPT from OpenAI, Gemini from Google, Co-Pilot from Microsoft, and Claude from Anthropic. 
Although all are based on the same foundational architecture \cite{Polverini2023Howunderstandinglargelanguagem}, since their initial release GenAI tools have rapidly evolved, with new models and updates releasing almost weekly. 
These updates have rapidly expanded their abilities and accuracy, making it difficult to pin down their exact capabilities. 

However, new technology takes time to be adopted, and this adoption takes time to study. 
It is therefore unsurprising that the research on GenAI's impact on physics research and educational practice is thin. 
Even so, certain trends and tentative conclusions have begun to emerge. 
First, several researchers have examined how GenAI tools perform on standard physics assessments or assignments, from concept inventories to essay questions. 
These articles have found varying AI performance, especially with tasks that require more complex reasoning \cite{LopezSimo2024ChallengingChatGPT} or visual interpretation \cite{polverini_performance_2025, polverini_performance_2024}. 
Some researchers have pointed out that the mistakes that GenAI tools make are reminiscent of the well-documented misconceptions students may have about different physics concepts \cite{Wheeler2023ChatGPTmisconceptions}. 
However, the most recent models seem to be adept enough to pass most physics exams and perform at least adequately at most tasks \cite{Pimbblet2025CanChatGPTpassaphysicsdegreeMa,Yeadon2024EvaluatingAIandhumanauthorship,polverini_performance_2024}, in many cases better than the standard physics student \cite{tschisgale_evaluating_2025, kortemeyer_multilingual_2025}. 
Physics students may benefit from using GenAI tools, for example in getting hints and guidance on how to solve problems \cite{Lu2025Incentivizingsupplementalmatha, Tong2025ExploringtheroleofhumanAIcolla, perez_linde_going_2025, avila_using_2024}, and supporting student interest and engagement in physics \cite{Lademann2025Augmentinglearningenvironments}. 

With regard to use of AI tools in physics (and science) teaching, several researchers have found that they have significant potential to help teachers in a variety of ways, including assessment \cite{kortemeyer_toward_2023,Kortemeyer2025AssessingconfidenceinAIassiste, Zhai2023AIandformativeassessmentThetra, chen_grading_2025, fussell_comparing_2025}, task development \cite{kuchemann_can_2023}, and giving students feedback on submitted work \cite{Wan2024ExploringgenerativeAIassistedf}. 
However, teacher views on GenAI tools vary significantly \cite{Karahan2023Usingvideoelicitationfocusgrou, Kim2025ScienceTeachersApproachestoArt, Garofalo2025ScienceTeacherPerceptionsofthe, wattanakasiwich_physics_2025}.
For example, some teachers view AI as a simple digital tool to help plan lessons, while others are concerned it encourages student cheating and lacks real human creativity.
These views are critical to whether teachers choose to use such tools or not. 
Physics education researchers may also benefit from using GenAI tools, since much of their work focuses on qualitatively analyzing student thinking and behavior through the modalities of text \cite{Wulff2024Physicslanguageandlanguageusei, kieser_educational_2023}.

Considering the transformative potential of this technology and the scarcity of research exploring actual use of AI tools in physics, there is a major gap in the literature around how physicists can (and do) use these tools in their work. 
This use will be mediated by their perceptions of the tools themselves, their usefulness, and their reliability. 
In order to explore this subject, we have adopted the theoretical framework of epistemic framing, which has a rich history in the field of physics education research for studying ways in which students (and teachers) perceive and use different ideas and knowledge-building tools.

\section{Theoretical Framework: Epistemic Framing}

Framing refers to the way in which people can take diverse views on situations or events. 
It is often described as how someone might answer the question ``What is going on here?'' that is to say a particular ``lens'' they take on a situation. 
For instance, a joke might be framed as humorous or as offensive, depending on the listener and context. 
Epistemic framing refers more specifically to the way in which people (especially students) perceive learning situations and tools: for example, different students might frame a physics problem as an opportunity for sensemaking \cite{odden_defining_2019,odden_sensemaking_2018} or as a task that requires them to simply produce an answer with minimal effort \cite{chen_epistemic_2013}. 
Physics Education Research has a rich literature on different framings and their effects on learning in environments like laboratory instruction \cite{sundstrom_instructing_2023}, assessments \cite{shar_student_2020}, upper-division courses \cite{chari_student_2019, dini_case_2017}, collaborative group work \cite{scherr_student_2009}, problem-solving \cite{bing_analyzing_2009}, and more.

Researchers have also studied instructor framing, and how it impacts teaching \cite{russ_inferring_2013}. 
This work has established that instructor or teacher framing can have a significant impact on student behavior and experience \cite{chari_student_2019, sundstrom_instructing_2023}. 
However, framing is also a critical factor in the uptake of educational innovations; for instance, instructors' views on and framing of the learning process is a critical factor in whether they choose to engage in traditional lecture-based teaching or active learning \cite{dancy_framework_2007}. 
It is therefore important at this juncture, where GenAI has been available for several years but is still being integrated into educational systems, to explore the ways in which instructors are framing it and how that framing is operationalized in the form of different uses of AI (which we refer to as practices).

\section{Methods}
Interviews are a common tool for probing student and instructor framing of learning processes \cite{bodin_mapping_2012, shar_student_2020, bing_analyzing_2009}, especially when those processes are new or developing. 
In the present study, interviews were held with twelve physics professors in the time period of late spring and early fall of 2024. 
The interview protocol consisted of questions on how the professors were using GenAI in their teaching, which experiences they have had, whether they had received feedback from students on this, how they use the technology in their own research, and their thoughts on future applications.
Interviews were audio-recorded, transcribed, and then anonymized by replacing names with pseudonyms which appear in this paper. 
There is no correlation between the gender of the pseudonyms and the actual genders of the professors.

Next, the transcribed interviews were inductively coded by the first author using a thematic analysis methodology \cite{braun_using_2006} to identify themes in GenAI perception and usage. 
After an initial round of coding, both authors discussed the emerging themes and grouped them into six overlapping epistemic frames, collectively comprised of 19 GenAI practices. 
These practices spanned both teaching and research, and included strategies for both GenAI use and critical evaluation. 
Each frame was further associated with specific \textit{warrants}, that is, articulated reasons that support that particular frame or view \cite{bing_analyzing_2009}.
Thereafter, the first author went back to the dataset and deductively re-coded the transcripts using the defined practices, frames, and warrants to ensure that these themes were a consistent and accurate representation of the data. 
This second round of coding resulted in minor revisions to names and descriptions for both practices and frames, as well as consolidation and removal of some practices. 
Finally, descriptions of frames and practices were compiled into a report.
The results of this analysis are presented below.

\section{Results: Physics Professors' framings and use of generative AI}

Table \ref{tab:ai_practices} summarizes the identified GenAI practices. 
\begin{figure}[h!]
 \centering
        \includegraphics[width=1.0\textwidth]{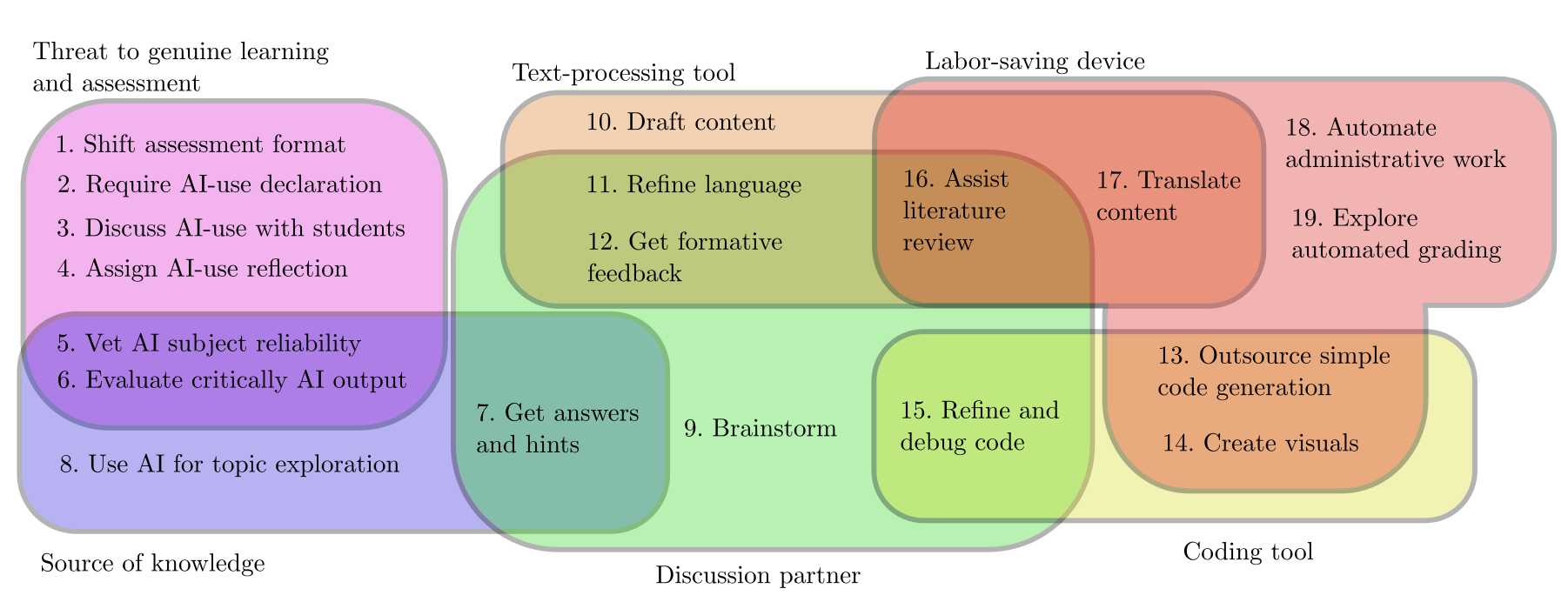}
 \caption{Organization of 19 GenAI practices within the six identified epistemic frames.}
\label{fig:epistemicframes}
\end{figure}
Table \ref{tab:epistemicframes} describes the identified frames, including the practices that they encompass. 
These frames and practices are also summarized in Figure \ref{fig:epistemicframes}.
\begin{table}[htbp]
    \centering
    \scriptsize{
    \begin{tabular}{p{3cm}p{10cm}}
        \toprule
        \textbf{Practice} & \textbf{Description} \\
        \midrule
        1. Shift assessment format & Altering or affirming assessment methods that mitigate the challenges GenAI poses to academic integrity. It often includes shifting from unsupervised written work to formats like oral exams or in-person tests without aids. \\
        \midrule
        2. Require AI-use declaration & Requiring students to disclose their use of GenAI in academic work, often in a dedicated section of their submission. \\
        \midrule
        3. Discuss AI-use with students & Engaging students in discussions about their learning process and the role of GenAI within it. \\
        \midrule
        4. Assign AI-use reflection & A specific form of promoting metacognition where students must submit a reflection note alongside their work. In this note, they describe their creative and analytical process, including how and why they used GenAI. \\
        \midrule
        5. Vet AI subject reliability & Professors personally testing the capabilities and accuracy of AI within their own subject area. It is often motivated by curiosity and the need to understand how students might use the tool for course-related questions. \\
        \midrule
        6. Evaluate critically AI output & Professors critically evaluating AI output or assigning students the task to do so themselves as part of their learning.  \\
        \midrule
        7. Get answers and hints & Using AI as a direct resource to obtain solutions to problems or to get a hint when stuck on a specific point. \\
        \midrule
        8. Use AI for topic exploration & Generating a broad overview of an unfamiliar topic. It serves as a starting point to quickly grasp the key concepts of a new field before consulting more detailed and reliable sources. \\
        \midrule
        9. Brainstorm & Using AI to generate a wide range of ideas or suggestions. The goal is not to get a correct answer, but to spark inspiration for a research question, a project idea, or a way to approach a problem. \\
        \midrule
        10. Draft content & Having GenAI generate entire sections of text, such as an introduction or a report, from a prompt. \\
        \midrule
        11. Refine language & Using GenAI to improve the grammar, style, and clarity of text that has already been written by a human. \\
        \midrule
        12. Get formative feedback & Getting feedback from GenAI on a written draft. This can involve asking general questions about clarity and logic, or checking the work against specific formal requirements. \\
        \midrule
        13. Outsource simple code generation & Delegating the writing of simple code to an AI. \\
        \midrule
        14. Create visuals & Using GenAI to generate code that produces visual representations of data or concepts. This includes creating plots, animations, or diagrams using languages like Python or specialized \LaTeX packages. \\
        \midrule
        15. Refine and debug code & Treating GenAI as an interactive assistant in the programming workflow. It includes asking for help with debugging, getting explanations of error messages, or getting feedback on code structure. \\
        \midrule
        16. Assist literature review & Using GenAI to manage the challenges of engaging with academic literature. It includes getting summaries of large amounts of text or asking for simplified explanations of technically difficult papers. \\
        \midrule
        17. Translate content & Using GenAI to translate material from one language to another. This applies both to human languages (e.g., Norwegian to English) and to programming languages (e.g., Matlab to Python). \\
        \midrule
        18. Automate administrative work & Using GenAI to handle routine, non-academic administrative tasks to save time. Examples include drafting formulaic emails, writing conference invitations, or re-formatting documents. \\
        \midrule
        19. Explore automated grading & Pondering the potential use of AI to assess student submissions automatically. It is mostly discussed as a future possibility for saving time on grading, particularly for more straightforward assignments. \\
        \bottomrule
    \end{tabular}
    }
    \caption{Physics professors' GenAI practices in teaching and research.}
    \label{tab:ai_practices}
\end{table}
\begin{table}[htbp]
    \begin{tabular}{p{2cm} p{6cm} p{4.5cm}}
        \toprule
        Frame & Evidence of frame (warrants) & Associated practices \\
        \midrule
        Threat to genuine learning and assessment & Focus on how AI disrupts current assessment practices; compare AI-use to cheating; discuss transparency in AI usage; problematize usage of AI to avoid learning.
        \newline
        Example: \textit{``There are surely some of those [students] we do not see [...] who use large language models to cheat'' (Jennifer).} 
        & 
        1. Shift assessment format \newline
        2. Require AI-use declaration \newline
        3. Discuss AI-use with students \newline
        4. Assign AI-use reflection \newline
        5. Vet AI subject reliability \newline
        6. Evaluate critically AI output  \\ \midrule
        Source of knowledge & Focus on AI's ability to solve physics problems, provide summaries of topics or explain hard concepts;
        \newline 
        Example: \textit{``Calculus is so well-trodden [...] [so] ChatGPT will almost always be right'' (Sarah).}  
        &
        5. Vet AI subject reliability \newline
        6. Evaluate critically AI output \newline
        7. Get answers and hints \newline
        8. Use AI for topic exploration 
        \\ \midrule
        Discussion partner & Describe AI as a ideation/brainstorming tool; Use words like ``discussing'' with the AI; \newline
        Example:  \textit{``Almost no [students] used AI like I had wanted. Namely as a Socratic partner in a discussion of a physics problem'' (Robert).}
        &
        7. Get answers and hints \newline
        9. Brainstorm \newline
        11. Refine language \newline
        12. Get formative feedback \newline
        15. Refine and debug code \newline
        16. Assist literature review
        \\ \midrule
        Coding tool & Focus on using AI in the process of writing computer code; \newline
        Example: \textit{``There is a lot of use [to ChatGPT]. Making a [...] draft for code. That is practical'' (Sarah).}
        &
        13. Outsource simple code generation \newline
        14. Create visuals \newline
        15. Refine and debug code 
        \\ \midrule
        Text-processing tool & Focus on AI's role in generating, refining, or summarizing text. \newline
        Example: \textit{``As a practical tool to write reports, it is a very good tool'' (John).} &
        10. Draft content \newline
        11. Refine language \newline
        12. Get formative feedback \newline
        16. Assist literature review \newline
        17. Translate content 
        \\ \midrule
        Labor-saving tool & Depict situations where AI can save time; \newline
        Example: \textit{``You can  feed drawings and Feynman-diagrams by hand, and then it will give the \LaTeX code as TikZ'' (John).} 
        &
        16. Assist literature review \newline
        17. Translate content \newline
        18. Automate administrative work \newline
        19. Explore automated grading
        \\
        \bottomrule
    \end{tabular}
    \caption{The identified epistemic frames, description of their warrants and the associated practices.}
    \label{tab:epistemicframes}
\end{table}
In what follows, we describe these frames in detail, contextualizing them with selected quotes from the interviews (translated into English) and referencing the relevant practices. 

\subsection{GenAI as a threat to genuine learning and assessment}

Powerful tools can be used in good ways and bad. 
This principle certainly applies to GenAI, and the most prevalent framing from the professors was a worry that GenAI can pose a threat to genuine learning and assessment in physics.
When viewing GenAI through this lens, the professors worried about students using it to bypass learning, replace writing, or offload their thinking.
In other words, as students adopted these practices, the professors feared that it would become hard to assess what the students had actually learned.
\begin{quote}
\textbf{David:} ``To test if the student has learned something is to ask a question, and then we expect an answer. And if I can get the answer automatically from a system [...] without having to think for myself, then it becomes difficult for us to judge what they have learned and what they have not.''
\end{quote}
David's focus on the ``system'' as an obstacle for ``judging what they have learned'' illustrates the theme of GenAI as a threat to genuine assessment.

To confront this threat, the professors emphasized the value of controlled assessment formats, for instance shifting away from unsupervised written tasks like take-home exams. 
Eight of the professors had either changed the assessment form in their course, or reaffirmed their current practice of using closed-form assessment.
For example, Jessica explained why she had switched to oral examinations in her course:
\begin{quote}
\textbf{Jessica:} ``We had AI and GPT in mind when we did that because [...] if people have to present orally and defend it [their ideas] orally, answer questions about their essay, then we have more assurance that they actually know what they are talking about.''
\end{quote}
Jessica's emphasis on needing assurance confirms that she views GenAI as a threat that requires a direct change in assessment.

Shifting assessment formats is the most direct way to address the challenge of GenAI and assessment, but the interviewed professors mentioned some alternatives. 
For instance, many of the professors chose instead to strengthen the integrity of existing open-form assessments. 
Eight of the professors had asked students to declare their use of AI in their projects, and five had openly discussed the pros and cons of using GenAI with students to promote metacognition.
All of the five professors that had engaged in this type of constructive dialogue said that students were appreciative of the openness around GenAI.
A few also promoted metacognition by assigning an AI-use reflection task along with written hand-in reports. 
Emily presented this as an effective way of gaining insight into student learning:
\begin{quote}
\textbf{Emily:} ``Those reflection notes were golden, actually. 
Because I gain a bit of insight into how they are thinking. [...] 
And they have to take an active stance as to what they have done.''
\end{quote}
By pointing to students having to take ``an active stance,'' Emily shows how assigning AI-use reflection promotes metacognition in learning.

This theme of AI as a threat permeated the discussions with the professors.
Warrants for it appeared roughly twice as often compared to warrants for any other frame.
This indicates that this particular frame was more of an overarching concern the professors held, which may have colored how the professors viewed the other possible uses. 
While we could have reflected this in Table \ref{tab:epistemicframes} and Figure \ref{fig:epistemicframes} by having this frame encompass all practices, we chose to include only the most relevant practices that were explicitly mentioned in response to the threat description.

\subsection{GenAI as a source of knowledge}

Since GenAI has been trained on a large repository of text data, it can answer questions about many fields, including physics. 
This capability allows GenAI to be a potential source of knowledge of physics, especially for students. 
Knowing this, eleven of the professors had spent time gauging the veracity of its physics ``knowledge,'' for example by asking questions from their own sub-fields to vet its subject reliability. 
Ashley noted that the output was often impressive: 
\begin{quote}
\textbf{Ashley:} ``It looks very impressive [...] it looks like it has some kind of grasp of common physics concepts like conservation of energy.'' 
\end{quote}
In this way, the professors were framing GenAI as an information repository, similar to a textbook or a knowledgeable colleague. 

At the same time, physicists are trained to critically examine sources of knowledge (leading, for example, to the common practice of posing critical questions to presenters at conferences).
This type of critical evaluation revealed that GenAI sometimes produced unfactual answers.
Ten out of the twelve professors interviewed touched upon the importance of critically evaluating GenAI output.
In the words of James, 
\begin{quote}
\textbf{James:}``I don't dare to trust anything [the AI says].''
\end{quote}
The distrust James expresses here was common and led many professors to say that using GenAI tools requires caution, unless one has a good way to verify its claims. 

Seven of the professors had extended this idea of critically evaluating GenAI output by assigning this kind of critical evaluation as an activity to their students. 
In this way, they embraced the challenge of untrustworthiness by shifting the learning objective from \textit{finding the answer} to \textit{evaluating the answer}. 
One professor, William, operationalized this approach by designing exam questions wherein students were presented with an AI-generated solution and tasked with finding errors. 
He explained the rationale for the task as follows:
\begin{quote}
\textbf{William:}``
So I thought, here we kill two birds with one stone! [...]
You can't always trust what you get presented [as a solution], so you have to actually know the physics.
[...]
[And] maybe also as a warning [about ChatGPT] for the students, that we have to think for ourselves. 
'' 
\end{quote}
This practice reframed the students' role into that of an expert critic or peer reviewer, implicitly training them in the same critical habits of mind demonstrated by the professors themselves. 
The two birds to which William refers consist of this reframing, and also that the students learned something about not blindly trusting GenAI answers.

Even though the professors expressed some caution about the reliability of the output, five of them had used GenAI to explore new topics. 
The professors cited typical situations where they were required to learn about an unfamiliar part of physics, for example when serving as an external evaluator for someone else's master's students. 
This practice of topic exploration was seen as professionally useful:
\begin{quote}
\textbf{Robert:}``If I'm trying to create an example about batteries, I find it very good to use [AI] to dig a bit more and get it to... Simply use it to explain to me what's happening.''
\end{quote}
Often, the professors would use this initial summary as a skeleton before digging into the details using more reliable sources.

\subsection{GenAI as a discussion partner}

To the uninitiated, conversations with GenAI can be strange: it often comes off as someone who is very helpful, surprisingly knowledgeable, extremely forgetful, and sometimes unjustifiably confident. 
All of these kinds of interactions, however, are predicated on the fact that we can ``speak'' with GenAI in our own language. 
This fact led eight of the professors to frame GenAI as a discussion partner, often using metaphors that described GenAI as a \textit{``buddy''} or a \textit{``sparring partner.''}
In this framing, the focus was not on how GenAI can provide answers, but rather how it can be used as a tool to discuss and refine one's own ideas:
\begin{quote}
    \textbf{Jessica: } ``There were several who said that yes, it was very useful for refining the research question.''
\end{quote}
Crafting and refining research questions is part of the creative work of a physicist, i.e., a form of brainstorming. 
Six of the professors had either engaged in brainstorming with GenAI or encouraged students to do so.
They saw opportunities for students too \textit{``bat around''} ideas or refine research questions on open-ended projects. 
Another practice they highlighted was using GenAI to give formative feedback, especially on writing; for instance, students might ask an AI agent for its \textit{``opinion''} on their writing to check for clarity and logical coherence.

The framing of GenAI as a discussion partner was mostly optimistic and the professors considered the constant availability of GenAI as beneficial.
This was especially the case for students who may lack human discussion partners, for instance due to distance learning or social anxiety:
\begin{quote}
\textbf{John:} ``For those who have few conversation partners, this is an incredibly good tool. Because when they are alone and can't ask me or other fellow students, they can ask ChatGPT.''
\end{quote}
In this way, John and others framed GenAI as a useful support for students who might otherwise have difficulty finding opportunities to discuss their ideas. 

However, some professors noted a significant gap between this ideal and the reality of student practice:
\begin{quote}
    \textbf{Robert:} ``there was basically no one who used it the way I had wished.''
\end{quote}
While Robert had hoped that students would engage in ``Socratic dialogue'' and sensemaking with the AI, in reality, he saw that students mostly used it for low-level technical tasks. 
Robert attributed this disappointing behavior to a lack of good role models on how to interact with the AI. 
He was not the only one to point out limitations of using GenAI as a discussion partner; other professors highlighted issues such as it being too positive, or not being sufficiently context-aware.
These limitations led four of the professors to emphasize the value of human-to-human interaction, thereby framing GenAI as a useful supplement, rather than a replacement, for academic dialogue.

\subsection{GenAI as a coding tool}

Coding has been an integral part of physics for many decades, and the professors had noticed how proficient GenAI was in writing and debugging code.
Thus, many of them framed GenAI as a useful tool for writing, editing, and debugging computer code in both physics research and teaching. 
The professors mostly viewed these capabilities with a pragmatic and positive attitude, because for many physicists, coding is primarily a means to an end.
\begin{quote}
\textbf{David:}``For me, as a physicist, coding is our tool. [...] The point is not to learn coding. [...] It is a tool to read in data, to make plots, or to visualize data, in order to then be able to interpret them and say something about the mechanisms behind the processes.''
\end{quote}
David's statement that learning coding ``is not the point'' exemplifies an attitude that was prevalent among many of the professors.  
As a consequence, they engaged in practices such as outsourcing simple code generation to GenAI or creating visuals. 
Nine of the professors had experimented with GenAI for this purpose and seven had explicitly encouraged students to do so themselves.
For example, Jessica had instructed her students as follows:
\begin{quote}
\textbf{Jessica:} ``I don't want you to spend time struggling with Python, if you can rather think about the big picture''
\end{quote}
These examples demonstrate how the professors viewed GenAI as another tool in the physicists' toolbox, useful to expedite certain tasks and thereby make time to focus on ``real physics.''

A minority of the professors with a stronger computational background saw the potential for a deeper integration with GenAI as a partner in coding rather than a simple code-generation tool. 
They described using GenAI to help debug code, get explanations for cryptic error messages or get feedback on code structure. 
To them, coding is not just another tool in the box, but an integral part of their physicist identity, and some expressed worry that by outsourcing code generation to GenAI, the students would not learn coding as well as they otherwise might. 
\begin{quote}
\textbf{Michael:} ``I imagine that it is useful sometimes in life [...] to start from an empty file and write the basic structure from scratch.'' 
\end{quote}
This is an example of the first framing of GenAI as a threat to genuine learning coloring Michael's perspective of GenAI as a coding tool.

These contrasting perspectives demonstrate a general feature of how GenAI enters into the physics community: specialization.
Physicists who do not see computation as part of their identity can outsource coding to the GenAI tool, while physicists who do can engage more deeply with the AI as a partner. 
In this way, GenAI allows physicists to specialize further and nurture their own interests. 

\subsection{GenAI as a text-processing tool}

For many, the most striking feature of GenAI is its ability to produce and process large amounts of text. 
Thus, in this fifth identified frame, the professors positioned  GenAI as a tool for text processing and refinement.
All of the professors expressed this view in some parts of their interviews, where they focused on GenAI's ability to perform a range of text-based tasks, from simple grammatical correction to complex summarization and translation. 
The professors specifically highlighted GenAI’s proficiency with language, positively comparing it to existing tools like spell checkers or Google Translate:
\begin{quote}
\textbf{Amanda:} 
What is the difference if [...] someone struggles with English and wants to write it in their mother tongue and uses Google translate, takes the result and pastes it into their dissertation... Where is the difference if I instead write something and use a language model?
\end{quote}
Amanda's questioning of the distinction between a language model (GenAI technology) and the commonly-used Google Translate tool shows that she positioned GenAI as a useful tool for text-processing and translation.

Within this frame, however, the professors expressed a hierarchy of acceptability.
They distinguished between using GenAI for granular language tasks such as getting feedback or language refinement, which were seen as legitimate, as opposed to using it for initial drafting or formulation of ideas, which was viewed with skepticism. 
\begin{quote}
\textbf{Emily:} ``...I know that I'll never learn the comma rules, so it can very well fix the commas for me. [...] I don't want ChatGPT to come in and think for me, then I don't get to think.'' 
\end{quote}
Emily contrasts the acceptable detail-level tasks of ``fixing commas'' with the more comprehensive task of drafting, which she equates with thinking. 

While using GenAI to draft content was seen by some as problematic (as exemplified by Emily's quote), three of the professors also viewed it with some pragmatism.
\begin{quote}
    \textbf{David:} If you then use ChatGPT to write a third of an introduction to something, and then use that as inspiration, or as a first draft, and then iterate on it, then I think that's okay.
\end{quote}
David expresses the view that drafting content within a certain context (``as a first draft'') can be acceptable. 
This was also the view of some of the other professors, who admitted having used GenAI to write grant proposals, which they viewed as acceptable.  

Most professors saw AI-use for refining language and getting formative feedback as benign, particularly for non-native English speakers.
Many compared these practices to having a colleague proofread a document. 
Similarly, the practice of using GenAI to translate content between languages was seen as legitimate and useful. 
In her interview, Jennifer recounted how an LLM's translation from Bokmål to Nynorsk (Norway's two official written languages) was excellent:
\begin{quote}
    \textbf{Jennifer: } ``it was so well-written that I in some cases had to go back and change the Bokmål version as well.''
\end{quote}

\subsection{GenAI as a labor-saving device}

Physicists, like most people, like saving time and energy. 
As such, all of the professors at some point discussed GenAI as a possible labor-saving tool. 
This frame was characterized by professors describing GenAI in terms of its utility for increasing personal and professional efficiency. 
Specifically, professors highlighted situations where GenAI could be used to save time from ``boring tasks'' and thereby give them opportunities to do more meaningful work, such as research, student interaction, or course design. 
\begin{quote}
\textbf{Amanda:}``Things like that I think can be very useful, and time-saving. So that I can spend my time in more sensible ways than writing boring things.''
\end{quote}
Amanda's focus on time-saving exemplifies how she sees labor-saving potential in this technology. 

Professors highlighted labor-saving strategies across all three pillars of academic work: teaching, research, and administration. 
Teaching-wise, they frequently mentioned the possibility of automated grading and feedback. 
Although many professors were skeptical of GenAI's ability to accurately assess students, they saw the potential in using it to automate the more menial aspects of grading or preparing for oral exams.
\begin{quote}
    \textbf{James: } If we want a wider assessment of the students than having them sit for four hours and calculate analytically [...] then we need help because [...] Reading and giving feedback on reports is very time-demanding... Ah, I feel exhausted just thinking about it. 
\end{quote}
James contrasts the need for a broader assessment with the work required to assess reports, and concludes that GenAI could be the labor-saving device needed to improve assessment in physics education.
Eight of the professors hoped that GenAI could someday automate parts of the grading process.

Research-wise, several professors highlighted the potential for using GenAI to assist in literature reviews by sorting through vast numbers of papers, or to create visuals by generating complex \LaTeX code for figures. 
However, only three professors had actually tried using GenAI for literature reviews, with mostly unsatisfactory results: they reported that double-checking the GenAI's results took more time than it would have taken to do the review from scratch. 
Half of the interviewees had used GenAI to save time in creating visuals, mostly through writing code that produced a plot based on a data set, which they found very useful. 

Administratively, the professors had used GenAI to automate routine tasks such as writing generic emails and translating text.

\section{Discussion and Conclusion}

From the identified practices and epistemic frames, we see that the interviewed physics professors had simultaneous positive and negative framings of GenAI. 
On the one hand, the professors viewed GenAI through five overlapping positive epistemic frames (2-6), which positioned GenAI as a useful tool for off-loading ``grunt'' work across different applications within teaching, research, and administration. 
For instance, by outsourcing simple code generation, text production, and administrative work to GenAI, these physicists were able to reclaim time for activities that were viewed as more interesting, like analytical work. 

On the other hand, the framing of GenAI as a threat to genuine learning and assessment permeated the discussion of the other frames.
That is, despite their optimism, many of the professors saw ways in which AI may threaten the current state of physics education. 
It is well established that learning to do physics takes time and effort, and as these professors noted, over-reliance on GenAI has the potential to harm these learning experiences.
For instance, although GenAI is a convenient tool for improving the quality of writing, writing is itself a reflective practice in physics \cite{hoehn_framework_2020}, and offloading too much of this work risks short-circuiting those productive thought processes.

This conflict is not surprising, as powerful technologies come with both upsides and downsides. 
Incorporating powerful technologies is not new for physics, a field that has welcomed in technologies such as the calculator, computer, and symbolic mathematical software. 
Perhaps, then, AI will simply become the next tool to be incorporated into the physicist's toolbox.

However, GenAI tools are still developing and their adoption is still in its infancy. 
This might explain why many of the discussed uses of AI were at the level of small perturbations to the standard work of a university physicist, geared towards improving efficiency.

Based on this observation, we argue that it is still too early to clearly see what effects GenAI may have on the practices of physics research and teaching.
Although at present GenAI tools are mostly being incorporated into standard physics practice in fairly small ways that do not significantly change the role or nature of the work, this trend may change in the coming years. 

Several of the interviews hint at ways in which the physicists viewed GenAI as something that may eventually come to redefine what it means to ``do'' and teach physics, in much the same way that computation has fundamentally altered the field. 
For instance, several professors shared thoughts on outsourcing the laborious task of assessing student work to GenAI tools. 
This would be a fundamental shift in responsibilities, not just a small perturbation, as it would redefine the role of the physics professor from both teacher and evaluator towards more of an engineer of physics learning environments.
Another domain where GenAI may transform the discipline is the social and communal aspects of physics. 
At present, most physicists get ideas and feedback through discussion and collaboration with colleagues.
GenAI may affect this dynamic in two opposite ways.
On the one hand, by offloading aspects of the ``grunt work'' of physics, the professors will have even more time for these kinds of discussions.
On the other, in GenAI tools they now have a non-human discussion partner who is available at any time, for any purpose.
This constant availability may then end up reducing the amount of time physicists spend discussing ideas and methods with their colleagues.

In response to these threats, the interviewed professors were already beginning to shift their teaching practices, such as reconsidering assessment formats or requiring AI-use declaration. 
In the time since these interviews were conducted, it seems likely that this process of adjustment has continued.
As researchers interested in the effects of GenAI on physics learning, we suggest that it may be time to consider what kinds of AI-focused training could be most beneficial, both to professors and students, as well as what AI literacy might look like within the domain of physics.
This kind of guidance would be useful for helping the physics community to thoughtfully and critically adapt AI to its needs.
By identifying these different framings and practices, the present study aims to take a first step in this direction.

We note that, as a qualitative case study of professionals at a single institution, this study has several limitations. 
First, the results represent a snapshot of the environment of one university, at a particular instant within the larger-scale process of GenAI integration; we therefore view these results as more of an existence-proof than a generalizable framework at this time. 
Second, given the extraordinary speed with which this technology is evolving, both the technology and human use and adoption will almost certainly continue to change.
However, it also seems fair to assume that as these tools continue to evolve, the current uses and strengths we have observed will themselves continue; in other words, although we cannot predict how they will change or improve in the future, the uses that have already been productively established seem likely to continue.
Third, given the limited size of the interviewed sample, the identified list of practices is certainly not exhaustive. 
Surprisingly, however, many of the interviewed professors repeatedly described the same practices, suggesting that these frames are indeed recurring themes, which may resonate with professional physicists at other institutions.

In future work, we hope that researchers will continue to study and track how these kinds of framings evolve over time, since the ways in which professionals frame these tools both reflect and drive their use.
Such studies could be conducted across a range of different types of institutions (small universities, large universities, national laboratories, or even high schools) and focus on a range of different use-cases.
We also see a need for parallel studies of how physics students are using and framing GenAI tools.

As these tools, the discipline of physics, and the societies within which both are situated continue to evolve, we expect that new uses will continue to emerge.
Physics as a discipline will continue to change, but this work shows that physicists are ready and willing to meet this challenge.

\ack{This work was funded by the Norwegian Directorate for Higher Education and Skills (HK-dir), which supports the University of Oslo’s Center for Computing in Science Education.}






\pagestyle{plain}
\bibliography{refs}
\bibliographystyle{unsrt}

\end{document}